\documentclass[reprint,aps,prl,showpacs,superscriptaddress,floatfix,byrevtex]{revtex4-1}
\usepackage[english]{babel}  % Hyphenation and whatnot
\usepackage{color}           % use colours
\usepackage{dcolumn}         % multiple columns
\usepackage{graphicx}        % Including graphics in the document
\usepackage{hyperref}        % links to references
\usepackage{bm}              % bold symbol (math)
\usepackage{float}           % Better handling of floats
\usepackage{pdfpages}        % insert PDF pages
\usepackage{amsmath}         % Americal Math Soc. tools for math typing
\usepackage{amssymb}
\usepackage{amstext}
\usepackage{amsopn}
\usepackage{amsfonts}
\usepackage{amsxtra}
\begin{document}

\preprint{Draft --- not for distribution}

%
% The title and the list of authors
%
\title{Optical properties of the perfectly compensated semimetal WTe$_2$}
\author{C. C. Homes}
\email{homes@bnl.gov}
\affiliation{Condensed Matter Physics and Materials Science Department,
  Brookhaven National Laboratory, Upton, New York 11973, USA}%
\author{M. N. Ali}
\author{R. J. Cava}
\affiliation{Department of Chemistry, Princeton University, Princeton,
  New Jersey 08544, USA}%
\date{\today ; ver.~5}

%
% The abstract goes here
%
\begin{abstract}
The optical properties of layered tungsten ditelluride have been measured over
a wide temperature and frequency range for light polarized in the \emph{a-b} planes.
A striking low-frequency plasma edge develops in the reflectance at low temperature
where this material is a perfectly compensated semimetal.  The optical conductivity is
described using a two-Drude model which treats the electron and hole pockets as separate
electronic subsystems.  At low temperature, one scattering rate collapses by over two
orders of magnitude, while the other also undergoes a significant, but less dramatic,
decrease; both scattering rates appear to display the quadratic temperature dependence
expected for a Fermi liquid.
%
% As the temperature decreases, one scattering rate collapses by over two orders of magnitude,
% yielding $1/\tau_1 \simeq 0.3$~cm$^{-1}$ at 5~K.  While the other scattering rate also decreases,
% at low temperature it is considerably larger, $1/\tau_2 \simeq 35$~cm$^{-1}$.  Interestingly, .
%
First principles electronic structure calculations reveal that the low-lying optical excitations
are due to direct transitions between the bands associated with the electron and hole pockets.
\end{abstract}
%
%  PACS numbers
%  63.20.-e     Phonons in crystal lattices
%
%  71.20.Be 	Transition metals and alloys
%
%  72.15.Lh 	Relaxation times and mean free paths
%
%  74.          *** Superconductivity ***
%
%  74.25.Bt T   Thermodynamic properties
%  74.25.nd     Raman and optical spectroscopy
%  74.25.Gz     Optical properties
%  74.25.Ha     Magnetic properties
%  74.25.Nf     Response to electromagnetic fields
%  74.70.-b     Superconducting materials other than cuprates
%  74.70.Xa 	Pnictides and chalcogenides
%  74.72.Bk     Y-based cuprates
%
%  77.22.Ch     Permittivity (dielectric function)
%
%  78.          Optical properties, condensed-matter spectroscopy and other interactions
%               of radiation and particles with condensed matter
%
%  78.30.-j     Infrared and Raman spectra
%
%  89.75.Da Systems obeying scaling laws
%
\preprint{Version 3}
\maketitle

%
% The main body of the text
%
% Introduction
%
%\section{Introduction}
%

Materials containing tellurium display a wide range of electronic properties
that make them of interest for a variety of different applications, including photovoltaics
\cite{basol14}, thermoelectrics \cite{goldsmid14}, \emph{p}-type semiconductors \cite{petersen73},
topological insulators \cite{qi11}, and superconductors \cite{liu10}.
To add to this wide variety of phenomena it has recently been proposed that the layered
transition metal dichalcogenide WTe$_2$ is a perfectly compensated semimetal at low temperature
\cite{ali14}, i.e., the number of electrons and holes per unit volume is identical.
A simple two-band model predicts that the magnetoresistance will saturate at high fields,
unless the material is compensated, in which case the magnetoresistance will increase
quadratically with field \cite{fawcett63}.  The magnetoresistance of WTe$_2$ increases
in precisely such a fashion; moreover, it becomes extremely large at low temperature and
and shows no signs of saturation, even at very high fields \cite{ali14}.  Electronic structure
and transport studies indicate that there are electron and hole pockets in this material
\cite{kabashima66,augustin00}, and angle resolved photoemission studies show that
at low temperature they are almost exactly the same size \cite{pletikosic14,jiang15},
resulting in a perfectly compensated semimetal.
%
% zero field
%
The zero-field behavior is equally dramatic with a very large residual resistivity ratio
of $\rho(300\,{\rm K})/\rho(5\,{\rm K}) \simeq 300$, with $\rho(5\,{\rm K})
\simeq 2\times 10^{-6}$~$\Omega\,{\rm cm}$ \cite{ali14}.  It is expected that such a dramatic
change in the resistivity should produce striking changes in the optical properties of this
material.

%
% Figure 1
%
\begin{figure}[t]
%
% manuscript
%
%\vspace*{-0.5cm}%
%\centerline{\includegraphics[width=0.90\columnwidth]{figure1.eps}}%
\centerline{\includegraphics[width=0.95\columnwidth]{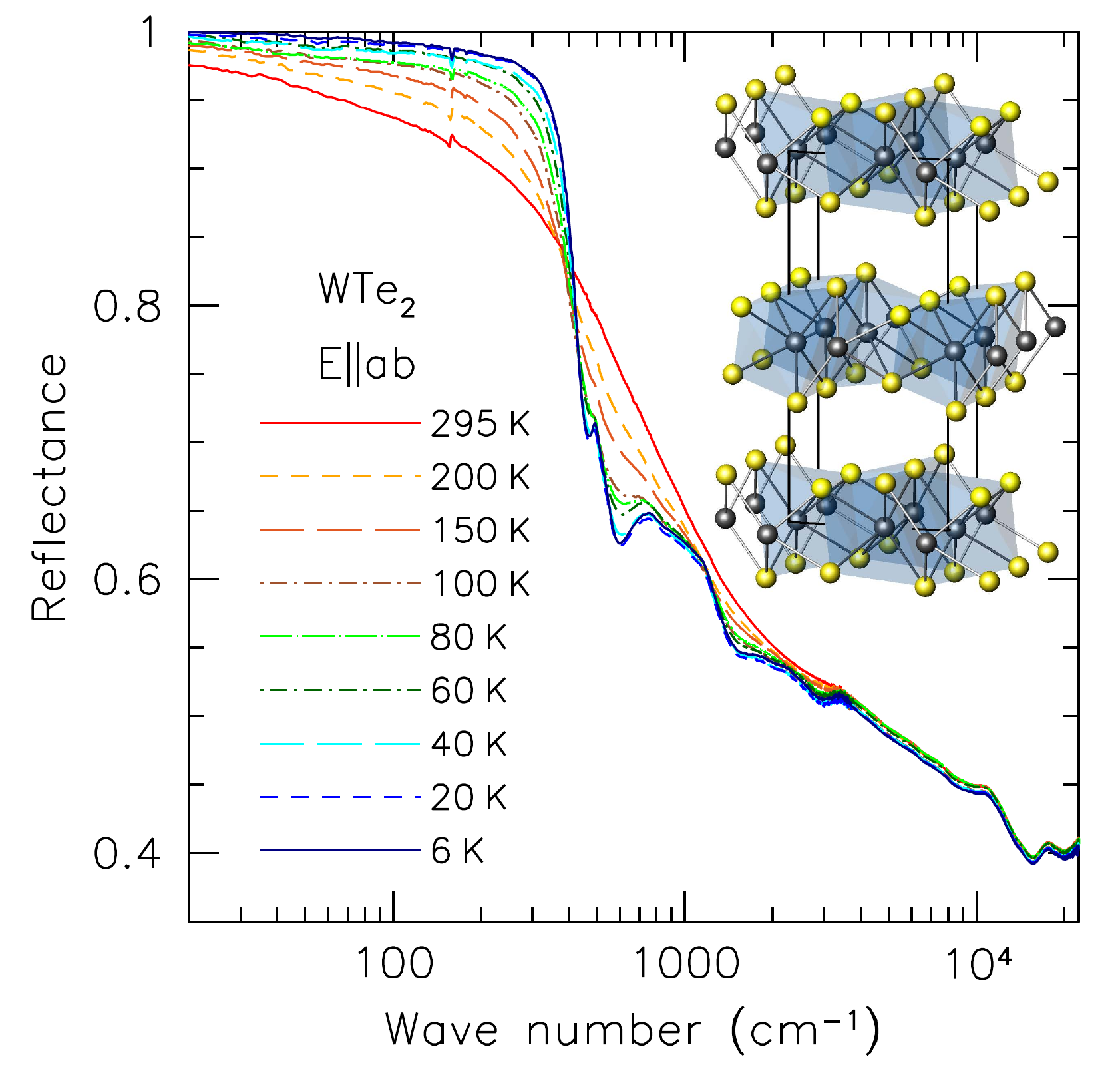}}%
%\vspace*{-1.0cm}%
\caption{(Color online) The temperature dependence of the reflectance of
in the infrared region of WTe$_2$ for light polarized in the \emph{a-b}
planes, revealing the formation of a sharp plasma edge in the reflectance
with decreasing temperature.
Inset: The unit cell is shown for the \emph{b-c} face projected along
the \emph{a} axis.}
% The tungsten atoms are surrounded by tellurium octahedra that are distorted
% due to the zig-zag nature of the tungsten chains.
%
\label{fig:reflec}
\end{figure}

%
% In this work...
%
In this study the optical properties of WTe$_2$ are determined in the \emph{a-b} plane
in the absence of a magnetic field over a wide frequency and temperature range.
The reflectance reveals the formation of a striking low-frequency plasma edge at
low temperature.  The complex conductivity indicates that the electron and hole pockets
give rise to carriers with dramatically different scattering rates.  While both follow
a quadratic temperature dependence and may be thought of as Fermi liquids, one scattering
rate decreases by more than two orders of magnitude at low temperature and is responsible
for the formation of a plasma edge in the reflectance.  Previous electronic structure
calculations are reproduced, showing electron and hole pockets of roughly equal size,
and extended to include the real part of the in-plane optical conductivity.  The
low-energy interband transitions are shown to originate from direct transitions
between the bands associated with the hole and electron pockets.

%
% Figure 2: optical conductivity
%
\begin{figure}[t]
%
% manuscript
%
%\vspace*{-0.5cm}%
%\centerline{\includegraphics[width=0.90\columnwidth]{figure2.eps}}%
\centerline{\includegraphics[width=0.95\columnwidth]{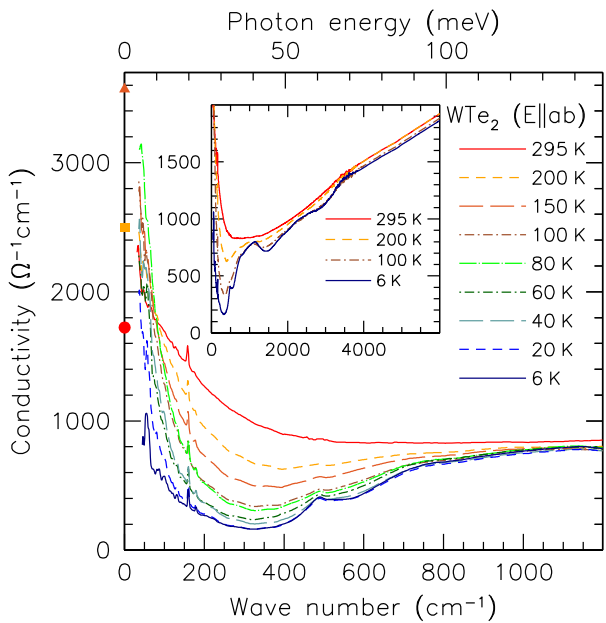}}%
%\vspace*{-1.0cm}%
\caption{(Color online) The temperature dependence of the real part of
the optical conductivity of WTe$_2$ for light polarized in the \emph{a-b}
planes.  The symbols along the $y$ axis denote the values for $\sigma_{\rm DC}$
obtained from transport measurements \cite{ali14}.
Inset: The optical conductivity for a restricted set of temperatures over
a wider frequency range.}
\label{fig:sigma}
\end{figure}

%
% Experiment
%
Single crystals of WTe$_2$ were grown using a bromine vapor transport method \cite{ali14}.
WTe$_2$ crystallizes in the orthorhombic $Pmn2_1$ space group \cite{mar92}, where the tungsten
atoms form chains along the \emph{a} axis and lie between sheets of tellurium atoms, forming
the \emph{a-b} planes.  The sheets stack along the \emph{c} axis and are easily exfoliated
(inset of Fig.~\ref{fig:reflec}).  The foil-like platelets were glued to a metal backing
plate for support and then mounted on optically-black cones.  The temperature dependence
of the reflectance, shown in Fig.~\ref{fig:reflec}, was measured for light polarized in the
\emph{a-b} planes over a wide frequency range ($\simeq 2$~meV to 3~eV) using an \emph{in situ}
overcoating technique \cite{homes93}.
The low-frequency reflectance is characteristic of a metallic response, increasing dramatically
with decreasing temperature and revealing a sharp plasma edge at $\sim 300 - 400$~cm$^{-1}$.
%
%this signals the collapse of a free-carrier scattering rate.  Other structure in the reflectance
%between $\sim 700 - 1300$~cm$^{-1}$ is attributed to low-energy interband transitions.
%
The optical conductivity has been determined from a Kramers-Kronig analysis of the reflectance
\cite{dressel-book}, which requires that the reflectance be supplied over the entire
frequency range; below the lowest measured frequency a metallic Hagen-Rubens form has been
employed, $1-R(\omega) \propto \sqrt{\omega}$, while above the highest-measured frequency
the reflectance is assumed to decrease as $1/\omega$ up to $1.5 \times 10^5$~cm$^{-1}$,
above which a free-electron $1/\omega^4$ extrapolation was used \cite{wooten}.

%
% Results - optical conductivity
%
The optical conductivity in the infrared region is shown in Fig.~\ref{fig:sigma}. At
high temperatures the response may be reasonably described by a single Drude component
(a Lorentzian centered at zero frequency where the full width at half maximum is the
free-carrier scattering rate), that gives way to a mid-infrared response that is
dominated by interband transitions.  As the temperature is reduced this Drude-like
feature narrows considerably leading to a transfer of spectral weight (area under the
conductivity curve) from high to low frequency, revealing several interband transitions,
shown more clearly in the inset of Fig.~\ref{fig:sigma}.  Above 2000~cm$^{-1}$ (0.25~eV)
the optical conductivity increases linearly with frequency until roughly 1~eV; we argue
that this linear behavior is due to the superposition of several different interband
transitions and not due to a strong renormalization of the free-carrier scattering
rate \cite{allen77,puchkov96}.
At high temperature there is reasonable agreement with $\sigma_1(\omega\rightarrow 0)$ and
the transport values for $\sigma_{\rm DC}=1/\rho$.  However, at low temperatures a
single Drude component is not capable of describing both the extremely high values for
$\sigma_{\rm DC}$ and the shape of the low-frequency conductivity.  Given that
there are both electron and hole pockets in this material, it is appropriate to
employ a model that accounts for both types of carriers \cite{wu10}.  The
complex dielectric function $\tilde\epsilon=\epsilon_1+i\epsilon_2$ of the
so-called two-Drude model is
\begin{equation}
  \tilde\epsilon(\omega) = \epsilon_\infty
    - \sum_{j=1}^2 {{\omega_{p,j}^2}\over{\omega^2+i\omega/\tau_{j}}}
    + \sum_k {{\Omega_k^2}\over{\omega_k^2 - \omega^2 - i\omega\gamma_k}},
\end{equation}
where $\epsilon_\infty$ is the real part at high frequency.  In the first sum
$\omega_{p,j}^2 = 4\pi n_je^2/m^\ast_j$ and $1/\tau_{j}$ are the square of the
plasma frequency and scattering rate for the delocalized (Drude) carriers in the
$j$th band, respectively, and $n_j$ and $m^\ast_j$ are the carrier concentration
and effective mass.  In the second summation, $\omega_k$, $\gamma_k$ and $\Omega_k$
are the position, width, and strength of the $k$th vibration or bound excitation.
The complex conductivity is $\tilde\sigma(\omega) = \sigma_1 +i\sigma_2 = -2\pi i \,\omega
[\tilde\epsilon(\omega) - \epsilon_\infty ]/Z_0$, where $Z_0 \simeq 377$~$\Omega$ is
the impedance of free space, yielding units for the conductivity of $\Omega^{-1}$cm$^{-1}$.

%
% Figure 3
%
\begin{figure}[t]
%
% manuscript
%
%\vspace*{-0.5cm}%
%\centerline{\includegraphics[width=0.90\columnwidth]{figure3.eps}}%
\centerline{\includegraphics[width=0.95\columnwidth]{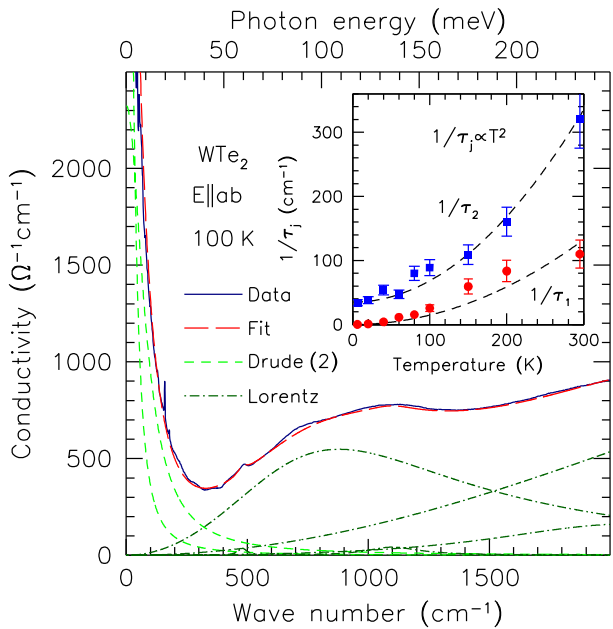}}%
%\vspace*{-1.0cm}%
\caption{(Color online) The fit to the complex optical conductivity of WTe$_2$
at 100~K.  The fit is composed of two Drude responses (dashed lines)
and several Lorentzian features (dot-dash lines); the linear combination (long-dash
line) reproduces the real part of the optical conductivity quite well.
Inset: The fitted values for $1/\tau_1$ (circles) and $1/\tau_2$ (squares);
the dashed lines are drawn as a guide to the eye.}
% the open symbols indicate that the parameter has been held fixed, while
%the solid symbols indicate that the parameter has been fitted.}
%
\label{fig:fit100}
\end{figure}

%
% Results of the fit at 100 K
%
The two-Drude model is fit simultaneously to both $\sigma_1$ and $\sigma_2$ using
a non-linear least-squares method.  The fit to the data at 100~K shown in
Fig.~\ref{fig:fit100} indicates that the optical conductivity can be reproduced quite
well using two Drude components, in addition to a series of Lorentzian oscillators at
$\simeq 800$, 1100 and 1950~cm$^{-1}$ ($\simeq 100$, 136 and 240~meV); other bound
excitations above $\simeq 3500$~cm$^{-1}$ (430~meV) are also included in the total
fit but not discussed.  At 100~K the two Drude contributions are of roughly equal
weight with $\omega_{p,1} = 3000\pm 200$~cm$^{-1}$, $1/\tau_1 = 26\pm 5$~cm$^{-1}$
and $\omega_{p,2} = 3400\pm 200$~cm$^{-1}$, $1/\tau_2 = 89\pm 12$~cm$^{-1}$; fits
at 80 and 150~K returned similar values for the plasma frequencies, so they have
been assumed to be temperature independent and are not fitted parameters.
%
% At higher temperatures
%
While fitting to both scattering rates at 100~K works reasonably well, at higher
temperatures the broad nature of the conductivity leads to larger values for the
scattering rates and increasing uncertainties.  At lower temperatures the
dramatic increase in $\sigma_{\rm DC}$ below 100~K suggests that $1/\tau_1$ is
becoming too small to be reliably determined from fits to $\tilde\sigma(\omega)$.
In order to provide an anchor for the fits to the scattering rates, $1/\tau_1$ is
first estimated from the Drude formula $1/\tau_1 = 2\pi\,\omega_{p,1}^2/(Z_0\,\sigma_{0})$.
For those temperatures where a direct comparison between transport and optics
is possible, it appears that $\sigma_0 = \sigma_1(\omega\rightarrow 0) \simeq
1.2\,\sigma_{\rm DC}$, so we have used the approximation $\sigma_0 \simeq 1.2\,\sigma_{\rm DC}$.
Treating $1/\tau_1$ as static, only $1/\tau_2$ is fit to the optical conductivity,
allowing the contribution $\sigma_{02} = 2\pi\,\omega_{p,2}^2\,\tau_2/Z_0$ to be
calculated where $\sigma_0=\sigma_{01} + \sigma_{02}$; the iterative improvement
in $1/\tau_1$ is determined from $1/\tau_1 = 2\pi\,\omega_{p,1}^2/(Z_0\,\sigma_{01})$.
The temperature dependence of the transport values of $\rho$, $\sigma_0$, and
the scattering rates determined in this fashion are summarized in Table~\ref{tab:results}.
The temperature dependence of $1/\tau_1$ and $1/\tau_2$ is shown in the inset of
Fig.~\ref{fig:fit100} as circles and squares, respectively.   While there is some
uncertainty of the scattering rates at high temperature, in both cases they follow
a roughly quadratic temperature dependence, suggesting both sets of carriers
resemble a Fermi liquid \cite{ghurzi59,maslov12,berthod13}.  While $1/\tau_2$ decreases
by roughly an order of magnitude at low temperature, it does not rival the
dramatic collapse of $1/\tau_1$ from $\simeq 110$~cm$^{-1}$ at 295~K to
$\simeq 0.3$~cm$^{-1}$ at 5~K, associated with the formation of the plasma edge
in the reflectance.
At low temperature the extremely large, non-saturating magnetoresistance indicates
that this material is perfectly compensated \cite{pletikosic14}, \emph{i.e.} $n_1 = n_2$.
From the values of $\omega_{p,1}$ and $\omega_{p,2}$, it may be shown that carrier masses
are similar, $m_2 \simeq 0.8 m_1$, in good agreement with Shubnikov-de-Hass oscillation
measurements \cite{cai15}.

%
% Phonons and interband transitions
%
%While the free-carriers dominate the low-frequency optical conductivity in
%Fig.~\ref{fig:sigma}, there are two sharp features at $\simeq 159$ and 177~cm$^{-1}$.
%The irreducible vibrational representation for the infrared-active modes is \cite{kong15}
%
%$$
%  \Gamma_{\rm IR} = 11\,A_1 + 5\,B_1 + 11\,B_2,
%$$
%which are active along the $c$, $b$ and $a$ axes, respectively; in addition, all of
%these modes are expected to be Raman active.  While a substantial number of infrared-active
%modes are expected, a simple empirical force-constant model \cite{vibratz} indicates
%that only two in-plane modes are expected to display any significant infrared intensity;
%this is consistent with the fact that only two vibrations are strong enough to be observed
%in Fig.~\ref{fig:sigma}.  The vibrations at 159 and 177~cm$^{-1}$ are tentatively assigned
%to $B_1$ and $B_2$ modes, respectively.  A rather unusual feature at $\simeq 480$~cm$^{-1}$
%(60~meV) emerges at low temperature; however, since this is well above the highest
%calculated phonon frequency \cite{kong15} it is not a lattice vibration.

%%%%%%%%%%%%%%%%%%%%%%%%%%%%%%%%%%%%%%%%%%%%%%%%%%%%%%%%%%%%%%%%%%%%%%%%%%%%%%%
%
% Table I - Drude-Lorentz parameter fits
%
\begin{table}[tb]
\caption{Transport values $\rho$ and $\sigma_{0}$, and optical scattering rates
$1/\tau_1$ and $1/\tau_2$.}
%
% The transport values for (taken from ) and the measured and estimated values for $\sigma_{0}
% = \sigma_1(\omega\rightarrow 0)$ for WTe$_2$. The scattering rates  are determined using an
% iterative process and the two-Drude model.}
%
\begin{ruledtabular}
\begin{tabular}{c cc cc}

  T & $\rho^{\rm a}$ & $\sigma_{0}$ & $1/\tau_1$ & $1/\tau_2$ \\
  (K) & ($\mu\Omega\,{\rm cm}$) & ($\times 10^3\,\Omega^{-1}{\rm cm}^{-1}$) & (cm$^{-1}$) & (cm$^{-1}$) \\
\cline{1-5}
   295 &  580 & 2.07 & $110\pm 22$  & $320\pm 45$ \\
   200 &  400 & 3.00 &  $84\pm 16$  & $160\pm 22$ \\
   150 &  280 & 4.28 &  $60\pm 12$  & $109\pm 15$ \\
   100 &  150 & 8.00 &  $26\pm  5$  &  $89\pm 12$ \\
    80 &  100 & 12.0 &  $16\pm  3$  &  $80\pm 11$ \\
    60 &   70 & 17.2 & $12\pm 2$  &  $47\pm 6$ \\
    40 &   30 & 40.0 & $4.1\pm 0.8$   &  $54\pm 7$ \\
    20 &    7 &  171 & $0.90\pm 0.2$ &  $38\pm 5$ \\
     6 &    2 &  600 & $0.25\pm 0.05$ &  $34\pm 4$ \\
\end{tabular}
\end{ruledtabular}
\footnotetext[1] {Ref.~\onlinecite{ali14}}
\label{tab:results}
%
%\vspace*{-0.5cm}
%
\end{table}
%%%%%%%%%%%%%%%%%%%%%%%%%%%%%%%%%%%%%%%%%%%%%%%%%%%%%%%%%%%%%%%%%%%%%%%%%%%%%%%
%

%
% Interband transitions
%
There are a series of excitations in the infrared at $\simeq 100, 136$ and
240~meV, as well as several others above 0.4~eV; these absorptions are in general
quite broad and have a substantial amount of spectral weight at low frequency.
%
%The band structure of this material indicates that there are numerous bands close
%the Fermi level that are candidates for low-energy interband transitions. Initial
%
Previous band structure studies of WTe$_2$ revealed a semimetallic character with
small electron and hole pockets \cite{ali14,augustin00}; these results have been
reproduced and extended in this work (the details of the calculations
\cite{singh,singh91,wien2k} are discussed in the supplementary materials).
%
% a more recent study \cite{ali14} revealed a low-temperature Fermi surface of electron
% and hole pockets displaced slightly from the zone center along the chain direction.
% The results of Ali \emph{et al.} \cite{ali14} have been reproduced and extended in
% this work; the details and results of these calculations \cite{singh,singh91,wien2k}
% are discussed and shown in the supplementary materials.
%
% this work with the local spin density approximation (LSDA) using the full-potential
% linearized augmented plane-wave (FP-LAPW) method \cite{singh} with local-orbital
% extensions \cite{singh91} in the WIEN2k implementation \cite{wien2k};
%

%
% Figure 4
%
\begin{figure}[t]
%
% manuscript
%
%\vspace*{-0.5cm}%
%\centerline{\includegraphics[width=0.90\columnwidth]{figure4.eps}}%
\centerline{\includegraphics[width=0.95\columnwidth]{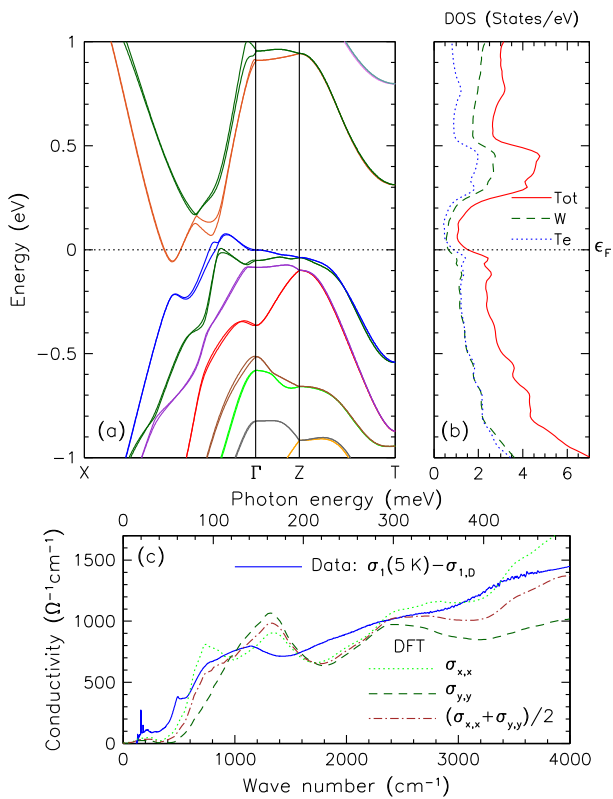}}%
%\vspace*{-1.0cm}%
\caption{(Color online) (a) The calculated electronic structure WTe$_2$
including the effects of spin-orbit coupling between several different
high symmetry points of the orthorhombic Brillouin zone.
(b) The total density of states (DOS), including the contributions from the
W and Te atoms.
(c) The real part of the calculated optical conductivity $\sigma_{x,x}$, $\sigma_{y,y}$
and average of the two, compared to the measured optical conductivity at 5~K with the
Drude terms removed.}
\label{fig:dft}
\end{figure}

The calculated electronic band structure including the effects of spin-orbit coupling
is shown in Fig.~\ref{fig:dft}(a) for paths between several different high-symmetry
points in the orthorhombic Brillouin zone, revealing both electron and hole bands
in the $\Gamma - {\rm X}$ direction, in good agreement with previous results
\cite{augustin00,ali14}.  A more detailed examination of the band structure,
(supplementary material), reveals several bands that lie just above
and below the Fermi level $\epsilon_{\rm F}$ that may be responsible for the
low-energy interband transitions.  To explore this further the total density of
states (DOS) for the W and Te atoms is calculated using a fine $k$-point mesh
($10\,000$ $k$ points) and shown in
%
% ($10\,000$ $k$ points, yielding a $41\times 23\times 10$ mesh) and shown in
%
Fig.~\ref{fig:dft}(b).  The low value for the DOS at $\epsilon_{\rm F}$ is
consistent with the semimetallic nature of this material.  The prominent
increase in the DOS in the $200-500$~meV above $\epsilon_{\rm F}$
is larger than $50-150$~meV interval over which interband transitions
are observed in the optical conductivity.  However, the optical conductivity
is a reflection of the joint density of states (JDOS) rather than just the DOS.
The real part of the optical conductivity has been calculated \cite{draxl06} using
the same fine $k$-point mesh for $\sigma_{x,x}$ and $\sigma_{y,y}$ (\emph{a}
and \emph{b} axes, respectively), shown in Fig.~\ref{fig:dft}(c), and
is compared to the experimental value for the conductivity at 5~K with the Drude
contributions removed.  Both $\sigma_{x,x}$ and $\sigma_{y,y}$ have a very
weak absorption at about 25~meV with the onset of much stronger absorptions
occurring above about 50~meV, with $\sigma_{x,x}$ displaying prominent features
at $\simeq 80$ and 160~meV, while only the 160~meV peak is present $\sigma_{y,y}$;
above $\simeq 200$~meV the two conductivities continue to increase steadily with
frequency.  The average of the two curves reproduces the experimental data
reasonably well.  The details of the contributions to the optical conductivity,
(supplementary material), reveal that these two strong absorptions
at 80 and 160~meV originate from direct transitions between the bands associated
with the hole and electron pockets.
The feature at 60~meV, which occurs at too high an energy to be a lattice
vibration \cite{kong15}, remains unexplained; however, it has been shown that
there are very low-energy excitations in this material.  Because this feature
only emerges at low temperature where the material is perfectly compensated,
it is likely that it is due to a low-energy interband transition that is
sensitive upon the details of the band structure that may not captured in
this calculation.

%
% Summary
%
To conclude, the complex optical properties of the perfectly compensated
semimetal WTe$_2$ have been measured in the \emph{a-b} planes for a variety
of temperatures over a wide frequency range.  The electron and hole pockets
are described using the two-Drude model, revealing that as the temperature
is reduced one scattering rate collapses by over two orders of magnitude,
resulting in the formation of a striking plasma edge in the reflectance,
while the other scattering rate is considerably larger and does not decrease
by nearly the same amount.  Both sets of carriers may be described as Fermi liquids.
%
% Interestingly, the quadratic temperature dependence of the scattering rates
% suggests they may both be described as Fermi liquids.
%
%First principles calculations are in good agreement with previous works and show
%that the Fermi surface consists of small electron and hole pockets.
%
The calculated optical conductivity indicates that low-energy absorptions are due to
direct transitions between the bands associated with the hole and electron pockets.
The emergence of a narrow low-energy absorption at low temperature is thought
to be a low-lying interband transition related to subtle changes in the electronic
structure that result in a perfectly compensated semimetal.

%
% acknowledgments
%\vspace*{-2.0mm}
%
\begin{acknowledgements}
We would like to acknowledge illuminating discussions with T. Valla.
This work is supported by the Office of Science, U.S. Department of Energy
under Contract No. DE-SC0012704 and by the Army Research Office, grant
W911NF-12-1-0461.
\end{acknowledgements}

%
%%%%%%%%%%%%%%%%%%%%%%%%%%%%%%%%%%%%%%%%%%%%%%%%%%%%%%%%%%%%%%%%%%%%%%%%%%%%%%
%
% References
%
%\bibliographystyle{apsrev4-1}
%\bibliography{tellurides}

\begin{thebibliography}{27}%
\makeatletter
\providecommand \@ifxundefined [1]{%
 \@ifx{#1\undefined}
}%
\providecommand \@ifnum [1]{%
 \ifnum #1\expandafter \@firstoftwo
 \else \expandafter \@secondoftwo
 \fi
}%
\providecommand \@ifx [1]{%
 \ifx #1\expandafter \@firstoftwo
 \else \expandafter \@secondoftwo
 \fi
}%
\providecommand \natexlab [1]{#1}%
\providecommand \enquote  [1]{``#1''}%
\providecommand \bibnamefont  [1]{#1}%
\providecommand \bibfnamefont [1]{#1}%
\providecommand \citenamefont [1]{#1}%
\providecommand \href@noop [0]{\@secondoftwo}%
\providecommand \href [0]{\begingroup \@sanitize@url \@href}%
\providecommand \@href[1]{\@@startlink{#1}\@@href}%
\providecommand \@@href[1]{\endgroup#1\@@endlink}%
\providecommand \@sanitize@url [0]{\catcode `\\12\catcode `\$12\catcode
  `\&12\catcode `\#12\catcode `\^12\catcode `\_12\catcode `\%12\relax}%
\providecommand \@@startlink[1]{}%
\providecommand \@@endlink[0]{}%
\providecommand \url  [0]{\begingroup\@sanitize@url \@url }%
\providecommand \@url [1]{\endgroup\@href {#1}{\urlprefix }}%
\providecommand \urlprefix  [0]{URL }%
\providecommand \Eprint [0]{\href }%
\providecommand \doibase [0]{http://dx.doi.org/}%
\providecommand \selectlanguage [0]{\@gobble}%
\providecommand \bibinfo  [0]{\@secondoftwo}%
\providecommand \bibfield  [0]{\@secondoftwo}%
\providecommand \translation [1]{[#1]}%
\providecommand \BibitemOpen [0]{}%
\providecommand \bibitemStop [0]{}%
\providecommand \bibitemNoStop [0]{.\EOS\space}%
\providecommand \EOS [0]{\spacefactor3000\relax}%
\providecommand \BibitemShut  [1]{\csname bibitem#1\endcsname}%
\let\auto@bib@innerbib\@empty
%</preamble>
\bibitem [{\citenamefont {Basol}\ and\ \citenamefont
  {McCandless}(2014)}]{basol14}%
  \BibitemOpen
  \bibfield  {author} {\bibinfo {author} {\bibfnamefont {B.~M.}\ \bibnamefont
  {Basol}}\ and\ \bibinfo {author} {\bibfnamefont {B.}~\bibnamefont
  {McCandless}},\ }\href {\doibase 10.1117/1.JPE.4.040996.} {\bibfield
  {journal} {\bibinfo  {journal} {J. Photon. Energy.}\ }\textbf {\bibinfo
  {volume} {4}},\ \bibinfo {pages} {040996} (\bibinfo {year}
  {2014})}\BibitemShut {NoStop}%
\bibitem [{\citenamefont {Goldsmid}(2015)}]{goldsmid14}%
  \BibitemOpen
  \bibfield  {author} {\bibinfo {author} {\bibfnamefont {H.~J.}\ \bibnamefont
  {Goldsmid}},\ }\href {\doibase 10.3390/ma7042577} {\bibfield  {journal}
  {\bibinfo  {journal} {Materials}\ }\textbf {\bibinfo {volume} {7}},\ \bibinfo
  {pages} {2577} (\bibinfo {year} {2015})}\BibitemShut {NoStop}%
\bibitem [{\citenamefont {Petersen}\ \emph {et~al.}(1973)\citenamefont
  {Petersen}, \citenamefont {Birkholz},\ and\ \citenamefont
  {Adler}}]{petersen73}%
  \BibitemOpen
  \bibfield  {author} {\bibinfo {author} {\bibfnamefont {K.~E.}\ \bibnamefont
  {Petersen}}, \bibinfo {author} {\bibfnamefont {U.}~\bibnamefont {Birkholz}},
  \ and\ \bibinfo {author} {\bibfnamefont {D.}~\bibnamefont {Adler}},\ }\href
  {\doibase 10.1103/PhysRevB.8.1453} {\bibfield  {journal} {\bibinfo  {journal}
  {Phys. Rev. B}\ }\textbf {\bibinfo {volume} {8}},\ \bibinfo {pages} {1453}
  (\bibinfo {year} {1973})}\BibitemShut {NoStop}%
\bibitem [{\citenamefont {Qi}\ and\ \citenamefont {Zhang}(2011)}]{qi11}%
  \BibitemOpen
  \bibfield  {author} {\bibinfo {author} {\bibfnamefont {X.-L.}\ \bibnamefont
  {Qi}}\ and\ \bibinfo {author} {\bibfnamefont {S.-C.}\ \bibnamefont {Zhang}},\
  }\href {\doibase 10.1103/RevModPhys.83.1057} {\bibfield  {journal} {\bibinfo
  {journal} {Rev. Mod. Phys.}\ }\textbf {\bibinfo {volume} {83}},\ \bibinfo
  {pages} {1057} (\bibinfo {year} {2011})}\BibitemShut {NoStop}%
\bibitem [{\citenamefont {Liu}\ \emph {et~al.}(2010)\citenamefont {Liu},
  \citenamefont {Hu}, \citenamefont {Qian}, \citenamefont {Fobes},
  \citenamefont {Mao}, \citenamefont {Bao}, \citenamefont {Reehuis},
  \citenamefont {Kimber}, \citenamefont {Proke\v{s}}, \citenamefont {Matas},
  \citenamefont {Argyriou}, \citenamefont {Hiess}, \citenamefont {Rotaru},
  \citenamefont {Pham}, \citenamefont {Spinu}, \citenamefont {Qiu},
  \citenamefont {Thampy}, \citenamefont {Savici}, \citenamefont {Rodriguez},\
  and\ \citenamefont {Broholm}}]{liu10}%
  \BibitemOpen
  \bibfield  {author} {\bibinfo {author} {\bibfnamefont {T.~J.}\ \bibnamefont
  {Liu}}, \bibinfo {author} {\bibfnamefont {J.}~\bibnamefont {Hu}}, \bibinfo
  {author} {\bibfnamefont {B.}~\bibnamefont {Qian}}, \bibinfo {author}
  {\bibfnamefont {D.}~\bibnamefont {Fobes}}, \bibinfo {author} {\bibfnamefont
  {Z.~Q.}\ \bibnamefont {Mao}}, \bibinfo {author} {\bibfnamefont
  {W.}~\bibnamefont {Bao}}, \bibinfo {author} {\bibfnamefont {M.}~\bibnamefont
  {Reehuis}}, \bibinfo {author} {\bibfnamefont {S.~A.~J.}\ \bibnamefont
  {Kimber}}, \bibinfo {author} {\bibfnamefont {K.}~\bibnamefont {Proke\v{s}}},
  \bibinfo {author} {\bibfnamefont {S.}~\bibnamefont {Matas}}, \bibinfo
  {author} {\bibfnamefont {D.~N.}\ \bibnamefont {Argyriou}}, \bibinfo {author}
  {\bibfnamefont {A.}~\bibnamefont {Hiess}}, \bibinfo {author} {\bibfnamefont
  {A.}~\bibnamefont {Rotaru}}, \bibinfo {author} {\bibfnamefont
  {H.}~\bibnamefont {Pham}}, \bibinfo {author} {\bibfnamefont {L.}~\bibnamefont
  {Spinu}}, \bibinfo {author} {\bibfnamefont {Y.}~\bibnamefont {Qiu}}, \bibinfo
  {author} {\bibfnamefont {V.}~\bibnamefont {Thampy}}, \bibinfo {author}
  {\bibfnamefont {A.~T.}\ \bibnamefont {Savici}}, \bibinfo {author}
  {\bibfnamefont {J.~A.}\ \bibnamefont {Rodriguez}}, \ and\ \bibinfo {author}
  {\bibfnamefont {C.}~\bibnamefont {Broholm}},\ }\href {\doibase
  10.1038/nmat2800} {\bibfield  {journal} {\bibinfo  {journal} {Nature Mater.}\
  }\textbf {\bibinfo {volume} {9}},\ \bibinfo {pages} {718} (\bibinfo {year}
  {2010})}\BibitemShut {NoStop}%
\bibitem [{\citenamefont {Ali}\ \emph {et~al.}(2014)\citenamefont {Ali},
  \citenamefont {Xiong}, \citenamefont {Flynn}, \citenamefont {Tao},
  \citenamefont {Gibson}, \citenamefont {Schoop}, \citenamefont {Liang},
  \citenamefont {Haldolaarachchige}, \citenamefont {Hirschberger},
  \citenamefont {Ong},\ and\ \citenamefont {Cava}}]{ali14}%
  \BibitemOpen
  \bibfield  {author} {\bibinfo {author} {\bibfnamefont {M.~N.}\ \bibnamefont
  {Ali}}, \bibinfo {author} {\bibfnamefont {J.}~\bibnamefont {Xiong}}, \bibinfo
  {author} {\bibfnamefont {S.}~\bibnamefont {Flynn}}, \bibinfo {author}
  {\bibfnamefont {J.}~\bibnamefont {Tao}}, \bibinfo {author} {\bibfnamefont
  {Q.~D.}\ \bibnamefont {Gibson}}, \bibinfo {author} {\bibfnamefont {L.~M.}\
  \bibnamefont {Schoop}}, \bibinfo {author} {\bibfnamefont {T.}~\bibnamefont
  {Liang}}, \bibinfo {author} {\bibfnamefont {N.}~\bibnamefont
  {Haldolaarachchige}}, \bibinfo {author} {\bibfnamefont {M.}~\bibnamefont
  {Hirschberger}}, \bibinfo {author} {\bibfnamefont {N.~P.}\ \bibnamefont
  {Ong}}, \ and\ \bibinfo {author} {\bibfnamefont {R.~J.}\ \bibnamefont
  {Cava}},\ }\href {\doibase 10.1038/nature13763} {\bibfield  {journal}
  {\bibinfo  {journal} {Nature (London)}\ }\textbf {\bibinfo {volume} {514}},\
  \bibinfo {pages} {205–208} (\bibinfo {year} {2014})}\BibitemShut {NoStop}%
\bibitem [{\citenamefont {Fawcett}\ and\ \citenamefont
  {Reed}(1963)}]{fawcett63}%
  \BibitemOpen
  \bibfield  {author} {\bibinfo {author} {\bibfnamefont {E.}~\bibnamefont
  {Fawcett}}\ and\ \bibinfo {author} {\bibfnamefont {W.~A.}\ \bibnamefont
  {Reed}},\ }\href {\doibase 10.1103/PhysRev.131.2463} {\bibfield  {journal}
  {\bibinfo  {journal} {Phys. Rev.}\ }\textbf {\bibinfo {volume} {131}},\
  \bibinfo {pages} {2463} (\bibinfo {year} {1963})}\BibitemShut {NoStop}%
\bibitem [{\citenamefont {Kabashima}(1966)}]{kabashima66}%
  \BibitemOpen
  \bibfield  {author} {\bibinfo {author} {\bibfnamefont {S.}~\bibnamefont
  {Kabashima}},\ }\href {\doibase 10.1143/JPSJ.21.945} {\bibfield  {journal}
  {\bibinfo  {journal} {J. Phys. Soc. Jpn.}\ }\textbf {\bibinfo {volume}
  {21}},\ \bibinfo {pages} {945} (\bibinfo {year} {1966})}\BibitemShut
  {NoStop}%
\bibitem [{\citenamefont {Augustin}\ \emph {et~al.}(2000)\citenamefont
  {Augustin}, \citenamefont {Eyert}, \citenamefont {B\"oker}, \citenamefont
  {Frentrup}, \citenamefont {Dwelk}, \citenamefont {Janowitz},\ and\
  \citenamefont {Manzke}}]{augustin00}%
  \BibitemOpen
  \bibfield  {author} {\bibinfo {author} {\bibfnamefont {J.}~\bibnamefont
  {Augustin}}, \bibinfo {author} {\bibfnamefont {V.}~\bibnamefont {Eyert}},
  \bibinfo {author} {\bibfnamefont {T.}~\bibnamefont {B\"oker}}, \bibinfo
  {author} {\bibfnamefont {W.}~\bibnamefont {Frentrup}}, \bibinfo {author}
  {\bibfnamefont {H.}~\bibnamefont {Dwelk}}, \bibinfo {author} {\bibfnamefont
  {C.}~\bibnamefont {Janowitz}}, \ and\ \bibinfo {author} {\bibfnamefont
  {R.}~\bibnamefont {Manzke}},\ }\href {\doibase 10.1103/PhysRevB.62.10812}
  {\bibfield  {journal} {\bibinfo  {journal} {Phys. Rev. B}\ }\textbf {\bibinfo
  {volume} {62}},\ \bibinfo {pages} {10812} (\bibinfo {year}
  {2000})}\BibitemShut {NoStop}%
\bibitem [{\citenamefont {Pletikosi{\'c}}\ \emph {et~al.}(2014)\citenamefont
  {Pletikosi{\'c}}, \citenamefont {Ali}, \citenamefont {Fedorov}, \citenamefont
  {Cava},\ and\ \citenamefont {Valla}}]{pletikosic14}%
  \BibitemOpen
  \bibfield  {author} {\bibinfo {author} {\bibfnamefont {I.}~\bibnamefont
  {Pletikosi{\'c}}}, \bibinfo {author} {\bibfnamefont {M.~N.}\ \bibnamefont
  {Ali}}, \bibinfo {author} {\bibfnamefont {A.~V.}\ \bibnamefont {Fedorov}},
  \bibinfo {author} {\bibfnamefont {R.~J.}\ \bibnamefont {Cava}}, \ and\
  \bibinfo {author} {\bibfnamefont {T.}~\bibnamefont {Valla}},\ }\href
  {\doibase 10.1103/PhysRevLett.113.216601} {\bibfield  {journal} {\bibinfo
  {journal} {Phys. Rev. Lett.}\ }\textbf {\bibinfo {volume} {113}},\ \bibinfo
  {pages} {216601} (\bibinfo {year} {2014})}\BibitemShut {NoStop}%
\bibitem [{\citenamefont {Jiang}\ \emph {et~al.}()\citenamefont {Jiang},
  \citenamefont {Tang}, \citenamefont {Pan}, \citenamefont {Liu}, \citenamefont
  {Niu}, \citenamefont {Wang}, \citenamefont {Xu}, \citenamefont {Yang},
  \citenamefont {Xie}, \citenamefont {Song}, \citenamefont {Wan},\ and\
  \citenamefont {Feng}}]{jiang15}%
  \BibitemOpen
  \bibfield  {author} {\bibinfo {author} {\bibfnamefont {J.}~\bibnamefont
  {Jiang}}, \bibinfo {author} {\bibfnamefont {F.}~\bibnamefont {Tang}},
  \bibinfo {author} {\bibfnamefont {X.~C.}\ \bibnamefont {Pan}}, \bibinfo
  {author} {\bibfnamefont {H.~M.}\ \bibnamefont {Liu}}, \bibinfo {author}
  {\bibfnamefont {X.~H.}\ \bibnamefont {Niu}}, \bibinfo {author} {\bibfnamefont
  {Y.~X.}\ \bibnamefont {Wang}}, \bibinfo {author} {\bibfnamefont {D.~F.}\
  \bibnamefont {Xu}}, \bibinfo {author} {\bibfnamefont {H.~F.}\ \bibnamefont
  {Yang}}, \bibinfo {author} {\bibfnamefont {B.~P.}\ \bibnamefont {Xie}},
  \bibinfo {author} {\bibfnamefont {F.~Q.}\ \bibnamefont {Song}}, \bibinfo
  {author} {\bibfnamefont {X.~G.}\ \bibnamefont {Wan}}, \ and\ \bibinfo
  {author} {\bibfnamefont {D.~L.}\ \bibnamefont {Feng}},\ }\href@noop {} {\
  }\Eprint {http://arxiv.org/abs/1503.01422} {arXiv:1503.01422 [cond-mat]}
  \BibitemShut {NoStop}%
\bibitem [{\citenamefont {Mar}\ \emph {et~al.}(1992)\citenamefont {Mar},
  \citenamefont {Jobic},\ and\ \citenamefont {Ibers}}]{mar92}%
  \BibitemOpen
  \bibfield  {author} {\bibinfo {author} {\bibfnamefont {A.}~\bibnamefont
  {Mar}}, \bibinfo {author} {\bibfnamefont {S.}~\bibnamefont {Jobic}}, \ and\
  \bibinfo {author} {\bibfnamefont {J.~A.}\ \bibnamefont {Ibers}},\ }\href@noop
  {} {\bibfield  {journal} {\bibinfo  {journal} {J. Am. Chem. Soc.}\ }\textbf
  {\bibinfo {volume} {114}},\ \bibinfo {pages} {8963} (\bibinfo {year}
  {1992})}\BibitemShut {NoStop}%
\bibitem [{\citenamefont {Homes}\ \emph {et~al.}(1993)\citenamefont {Homes},
  \citenamefont {Reedyk}, \citenamefont {Crandles},\ and\ \citenamefont
  {Timusk}}]{homes93}%
  \BibitemOpen
  \bibfield  {author} {\bibinfo {author} {\bibfnamefont {C.~C.}\ \bibnamefont
  {Homes}}, \bibinfo {author} {\bibfnamefont {M.}~\bibnamefont {Reedyk}},
  \bibinfo {author} {\bibfnamefont {D.~A.}\ \bibnamefont {Crandles}}, \ and\
  \bibinfo {author} {\bibfnamefont {T.}~\bibnamefont {Timusk}},\ }\href
  {\doibase 10.1364/AO.32.002976} {\bibfield  {journal} {\bibinfo  {journal}
  {Appl. Opt.}\ }\textbf {\bibinfo {volume} {32}},\ \bibinfo {pages} {2976}
  (\bibinfo {year} {1993})}\BibitemShut {NoStop}%
\bibitem [{\citenamefont {Dressel}\ and\ \citenamefont
  {Gr{\"u}ner}(2001)}]{dressel-book}%
  \BibitemOpen
  \bibfield  {author} {\bibinfo {author} {\bibfnamefont {M.}~\bibnamefont
  {Dressel}}\ and\ \bibinfo {author} {\bibfnamefont {G.}~\bibnamefont
  {Gr{\"u}ner}},\ }\href@noop {} {\emph {\bibinfo {title} {Electrodynamics of
  Solids}}}\ (\bibinfo  {publisher} {Cambridge University Press},\ \bibinfo
  {address} {Cambridge},\ \bibinfo {year} {2001})\BibitemShut {NoStop}%
\bibitem [{\citenamefont {Wooten}(1972)}]{wooten}%
  \BibitemOpen
  \bibfield  {author} {\bibinfo {author} {\bibfnamefont {F.}~\bibnamefont
  {Wooten}},\ }\href@noop {} {\emph {\bibinfo {title} {Optical Properties of
  Solids}}}\ (\bibinfo  {publisher} {Academic Press},\ \bibinfo {address} {New
  York},\ \bibinfo {year} {1972})\ pp.\ \bibinfo {pages} {244--250}\BibitemShut
  {NoStop}%
\bibitem [{\citenamefont {Allen}\ and\ \citenamefont
  {Mikkelsen}(1977)}]{allen77}%
  \BibitemOpen
  \bibfield  {author} {\bibinfo {author} {\bibfnamefont {J.~W.}\ \bibnamefont
  {Allen}}\ and\ \bibinfo {author} {\bibfnamefont {J.~C.}\ \bibnamefont
  {Mikkelsen}},\ }\href {\doibase 10.1103/PhysRevB.15.2952} {\bibfield
  {journal} {\bibinfo  {journal} {Phys. Rev. B}\ }\textbf {\bibinfo {volume}
  {15}},\ \bibinfo {pages} {2952} (\bibinfo {year} {1977})}\BibitemShut
  {NoStop}%
\bibitem [{\citenamefont {Puchkov}\ \emph {et~al.}(1996)\citenamefont
  {Puchkov}, \citenamefont {Basov},\ and\ \citenamefont {Timusk}}]{puchkov96}%
  \BibitemOpen
  \bibfield  {author} {\bibinfo {author} {\bibfnamefont {A.}~\bibnamefont
  {Puchkov}}, \bibinfo {author} {\bibfnamefont {D.~N.}\ \bibnamefont {Basov}},
  \ and\ \bibinfo {author} {\bibfnamefont {T.}~\bibnamefont {Timusk}},\ }\href
  {\doibase 10.1088/0953-8984/8/48/023} {\bibfield  {journal} {\bibinfo
  {journal} {J. Phys.: Condens. Matter}\ }\textbf {\bibinfo {volume} {8}},\
  \bibinfo {pages} {10049} (\bibinfo {year} {1996})}\BibitemShut {NoStop}%
\bibitem [{\citenamefont {Wu}\ \emph {et~al.}(2010)\citenamefont {Wu},
  \citenamefont {Bari{\v{s}}i{\'{c}}}, \citenamefont {Kallina}, \citenamefont
  {Faridian}, \citenamefont {Gorshunov}, \citenamefont {Drichko}, \citenamefont
  {Li}, \citenamefont {Lin}, \citenamefont {Cao}, \citenamefont {Xu},
  \citenamefont {Wang},\ and\ \citenamefont {Dressel}}]{wu10}%
  \BibitemOpen
  \bibfield  {author} {\bibinfo {author} {\bibfnamefont {D.}~\bibnamefont
  {Wu}}, \bibinfo {author} {\bibfnamefont {N.}~\bibnamefont
  {Bari{\v{s}}i{\'{c}}}}, \bibinfo {author} {\bibfnamefont {P.}~\bibnamefont
  {Kallina}}, \bibinfo {author} {\bibfnamefont {A.}~\bibnamefont {Faridian}},
  \bibinfo {author} {\bibfnamefont {B.}~\bibnamefont {Gorshunov}}, \bibinfo
  {author} {\bibfnamefont {N.}~\bibnamefont {Drichko}}, \bibinfo {author}
  {\bibfnamefont {L.~J.}\ \bibnamefont {Li}}, \bibinfo {author} {\bibfnamefont
  {X.}~\bibnamefont {Lin}}, \bibinfo {author} {\bibfnamefont {G.~H.}\
  \bibnamefont {Cao}}, \bibinfo {author} {\bibfnamefont {Z.~A.}\ \bibnamefont
  {Xu}}, \bibinfo {author} {\bibfnamefont {N.~L.}\ \bibnamefont {Wang}}, \ and\
  \bibinfo {author} {\bibfnamefont {M.}~\bibnamefont {Dressel}},\ }\href
  {\doibase 10.1103/PhysRevB.81.100512} {\bibfield  {journal} {\bibinfo
  {journal} {Phys. Rev. B}\ }\textbf {\bibinfo {volume} {81}},\ \bibinfo
  {pages} {100512(R)} (\bibinfo {year} {2010})}\BibitemShut {NoStop}%
\bibitem [{\citenamefont {Gurzhi}(1959)}]{ghurzi59}%
  \BibitemOpen
  \bibfield  {author} {\bibinfo {author} {\bibfnamefont {R.~N.}\ \bibnamefont
  {Gurzhi}},\ }\href@noop {} {\bibfield  {journal} {\bibinfo  {journal} {Sov.
  Phys. JETP}\ }\textbf {\bibinfo {volume} {8}},\ \bibinfo {pages} {673}
  (\bibinfo {year} {1959})}\BibitemShut {NoStop}%
\bibitem [{\citenamefont {Maslov}\ and\ \citenamefont
  {Chubukov}(2012)}]{maslov12}%
  \BibitemOpen
  \bibfield  {author} {\bibinfo {author} {\bibfnamefont {D.~L.}\ \bibnamefont
  {Maslov}}\ and\ \bibinfo {author} {\bibfnamefont {A.~V.}\ \bibnamefont
  {Chubukov}},\ }\href {\doibase 10.1103/PhysRevB.86.155137} {\bibfield
  {journal} {\bibinfo  {journal} {Phys. Rev. B}\ }\textbf {\bibinfo {volume}
  {86}},\ \bibinfo {pages} {155137} (\bibinfo {year} {2012})}\BibitemShut
  {NoStop}%
\bibitem [{\citenamefont {Berthod}\ \emph {et~al.}(2013)\citenamefont
  {Berthod}, \citenamefont {Mravlje}, \citenamefont {Deng}, \citenamefont
  {\ifmmode~\check{Z}\else \v{Z}\fi{}itko}, \citenamefont {van~der Marel},\
  and\ \citenamefont {Georges}}]{berthod13}%
  \BibitemOpen
  \bibfield  {author} {\bibinfo {author} {\bibfnamefont {C.}~\bibnamefont
  {Berthod}}, \bibinfo {author} {\bibfnamefont {J.}~\bibnamefont {Mravlje}},
  \bibinfo {author} {\bibfnamefont {X.}~\bibnamefont {Deng}}, \bibinfo {author}
  {\bibfnamefont {R.}~\bibnamefont {\ifmmode~\check{Z}\else \v{Z}\fi{}itko}},
  \bibinfo {author} {\bibfnamefont {D.}~\bibnamefont {van~der Marel}}, \ and\
  \bibinfo {author} {\bibfnamefont {A.}~\bibnamefont {Georges}},\ }\href
  {\doibase 10.1103/PhysRevB.87.115109} {\bibfield  {journal} {\bibinfo
  {journal} {Phys. Rev. B}\ }\textbf {\bibinfo {volume} {87}},\ \bibinfo
  {pages} {115109} (\bibinfo {year} {2013})}\BibitemShut {NoStop}%
\bibitem [{\citenamefont {Cai}\ \emph {et~al.}(2015)\citenamefont {Cai},
  \citenamefont {Hu}, \citenamefont {He}, \citenamefont {Pan}, \citenamefont
  {Hong}, \citenamefont {Zhang}, \citenamefont {Zhang}, \citenamefont {Wei},
  \citenamefont {Mao},\ and\ \citenamefont {Li}}]{cai15}%
  \BibitemOpen
  \bibfield  {author} {\bibinfo {author} {\bibfnamefont {P.~L.}\ \bibnamefont
  {Cai}}, \bibinfo {author} {\bibfnamefont {J.}~\bibnamefont {Hu}}, \bibinfo
  {author} {\bibfnamefont {L.~P.}\ \bibnamefont {He}}, \bibinfo {author}
  {\bibfnamefont {J.}~\bibnamefont {Pan}}, \bibinfo {author} {\bibfnamefont
  {X.~C.}\ \bibnamefont {Hong}}, \bibinfo {author} {\bibfnamefont
  {Z.}~\bibnamefont {Zhang}}, \bibinfo {author} {\bibfnamefont
  {J.}~\bibnamefont {Zhang}}, \bibinfo {author} {\bibfnamefont
  {J.}~\bibnamefont {Wei}}, \bibinfo {author} {\bibfnamefont {Z.~Q.}\
  \bibnamefont {Mao}}, \ and\ \bibinfo {author} {\bibfnamefont {S.~Y.}\
  \bibnamefont {Li}},\ }\href {\doibase 10.1103/PhysRevLett.115.057202}
  {\bibfield  {journal} {\bibinfo  {journal} {Phys. Rev. Lett.}\ }\textbf
  {\bibinfo {volume} {115}},\ \bibinfo {pages} {057202} (\bibinfo {year}
  {2015})}\BibitemShut {NoStop}%
\bibitem [{\citenamefont {Singh}(1994)}]{singh}%
  \BibitemOpen
  \bibfield  {author} {\bibinfo {author} {\bibfnamefont {D.~J.}\ \bibnamefont
  {Singh}},\ }\href@noop {} {\emph {\bibinfo {title} {Planewaves,
  Pseudopotentials and the LAPW method}}}\ (\bibinfo  {publisher} {Kluwer
  Adademic},\ \bibinfo {address} {Boston},\ \bibinfo {year} {1994})\BibitemShut
  {NoStop}%
\bibitem [{\citenamefont {Singh}(1991)}]{singh91}%
  \BibitemOpen
  \bibfield  {author} {\bibinfo {author} {\bibfnamefont {D.}~\bibnamefont
  {Singh}},\ }\href {\doibase 10.1103/PhysRevB.43.6388} {\bibfield  {journal}
  {\bibinfo  {journal} {Phys. Rev. B}\ }\textbf {\bibinfo {volume} {43}},\
  \bibinfo {pages} {6388} (\bibinfo {year} {1991})}\BibitemShut {NoStop}%
\bibitem [{wie()}]{wien2k}%
  \BibitemOpen
  \href@noop {} {}\bibinfo {note} {P. Blaha, K. Schwarz, G.~K.~H. Madsen, D.
  Kvasnicka and J. Luitz, WIEN2k, {\it An augmented plane wave plus local
  orbitals program for calculating crystal properties} (Techn.
  Universit{\"{a}}t Wien, Austria, 2001).}\BibitemShut {Stop}%
\bibitem [{\citenamefont {Ambrosch-Draxl}\ and\ \citenamefont
  {Sofo}(2006)}]{draxl06}%
  \BibitemOpen
  \bibfield  {author} {\bibinfo {author} {\bibfnamefont {C.}~\bibnamefont
  {Ambrosch-Draxl}}\ and\ \bibinfo {author} {\bibfnamefont {J.~O.}\
  \bibnamefont {Sofo}},\ }\href {\doibase
  http://dx.doi.org/10.1016/j.cpc.2006.03.005} {\bibfield  {journal} {\bibinfo
  {journal} {Comp. Phys. Commun.}\ }\textbf {\bibinfo {volume} {175}},\
  \bibinfo {pages} {1} (\bibinfo {year} {2006})}\BibitemShut {NoStop}%
\bibitem [{\citenamefont {Kong}\ \emph {et~al.}(2015)\citenamefont {Kong},
  \citenamefont {Wu}, \citenamefont {Richard}, \citenamefont {Lian},
  \citenamefont {Wang}, \citenamefont {Yang}, \citenamefont {Shi},\ and\
  \citenamefont {Ding}}]{kong15}%
  \BibitemOpen
  \bibfield  {author} {\bibinfo {author} {\bibfnamefont {W.-D.}\ \bibnamefont
  {Kong}}, \bibinfo {author} {\bibfnamefont {S.-F.}\ \bibnamefont {Wu}},
  \bibinfo {author} {\bibfnamefont {P.}~\bibnamefont {Richard}}, \bibinfo
  {author} {\bibfnamefont {C.-S.}\ \bibnamefont {Lian}}, \bibinfo {author}
  {\bibfnamefont {J.-T.}\ \bibnamefont {Wang}}, \bibinfo {author}
  {\bibfnamefont {C.-L.}\ \bibnamefont {Yang}}, \bibinfo {author}
  {\bibfnamefont {Y.-G.}\ \bibnamefont {Shi}}, \ and\ \bibinfo {author}
  {\bibfnamefont {H.}~\bibnamefont {Ding}},\ }\href {\doibase
  10.1063/1.4913680} {\bibfield  {journal} {\bibinfo  {journal} {Appl. Phys.
  Lett.}\ }\textbf {\bibinfo {volume} {106}},\ \bibinfo {pages} {081906}
  (\bibinfo {year} {2015})}\BibitemShut {NoStop}%
\end{thebibliography}
%

%merlin.mbs apsrev4-1.bst 2010-07-25 4.21a (PWD, AO, DPC) hacked
%Control: key (0)
%Control: author (8) initials jnrlst
%Control: editor formatted (1) identically to author
%Control: production of article title (-1) disabled
%Control: page (0) single
%Control: year (1) truncated
%Control: production of eprint (0) enabled
%

\newpage \ \\

\newpage
\includepdf{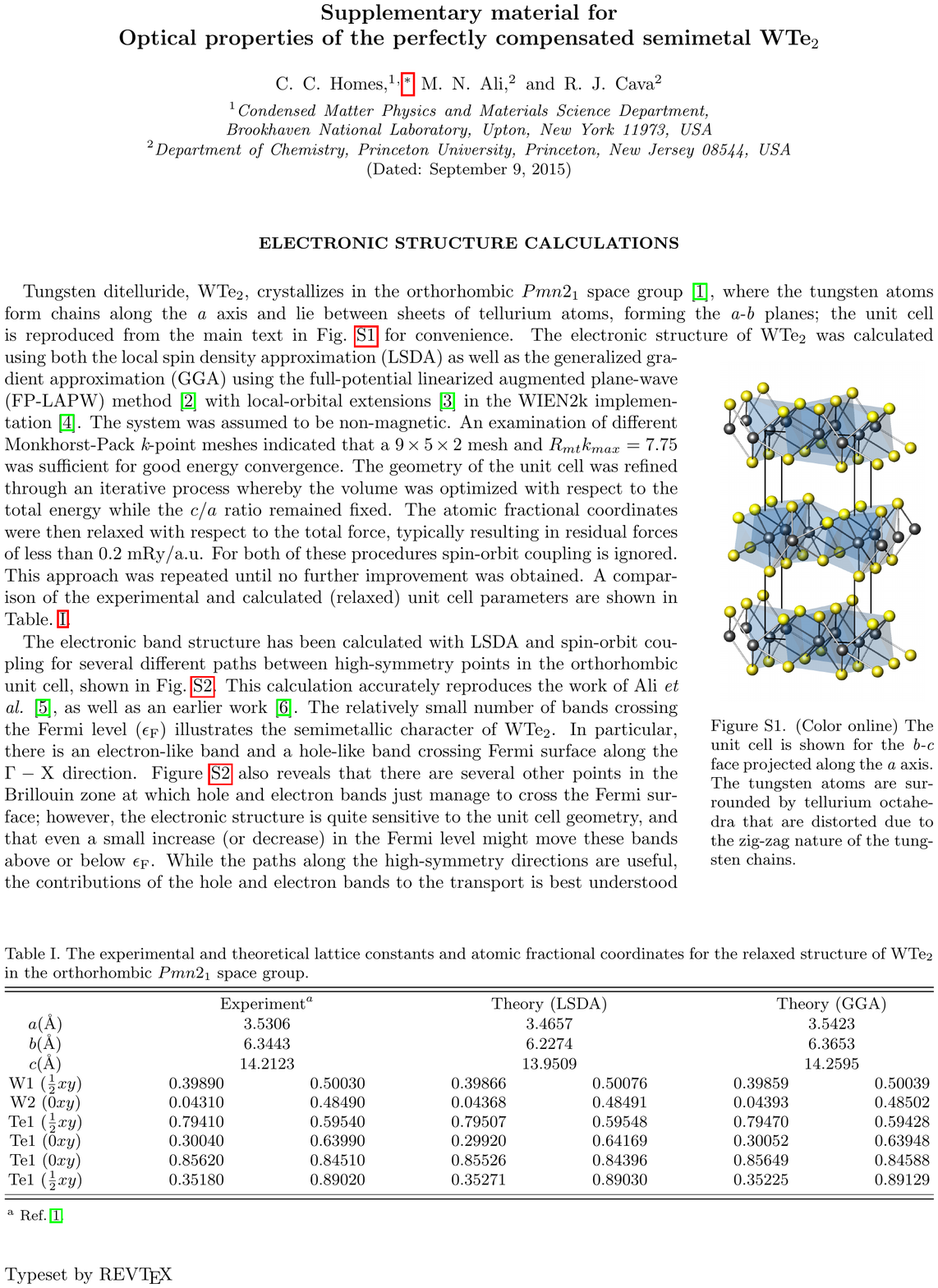} \ \\ % flush the page

\newpage
\includepdf{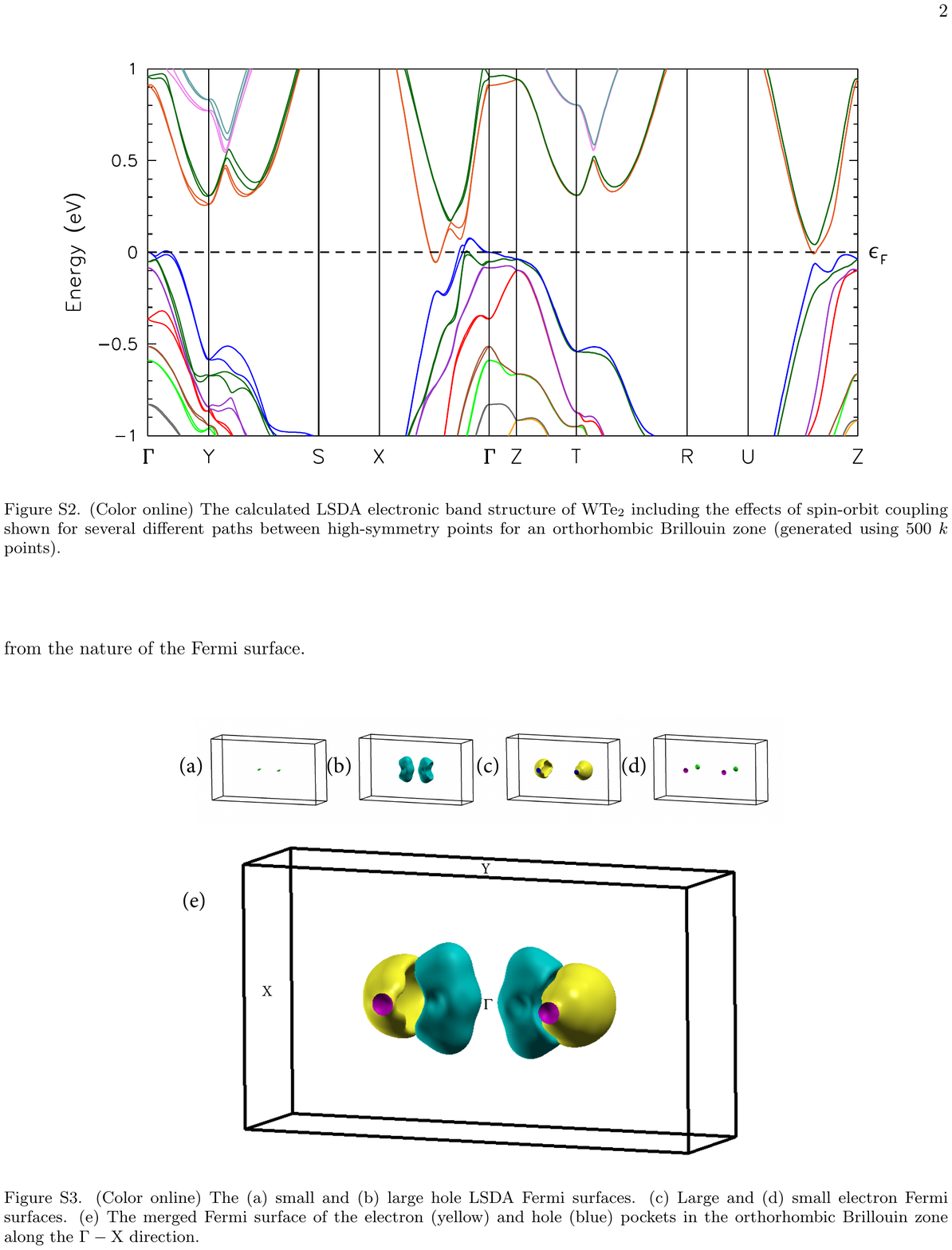} \ \\ % flush the page

\newpage
\includepdf{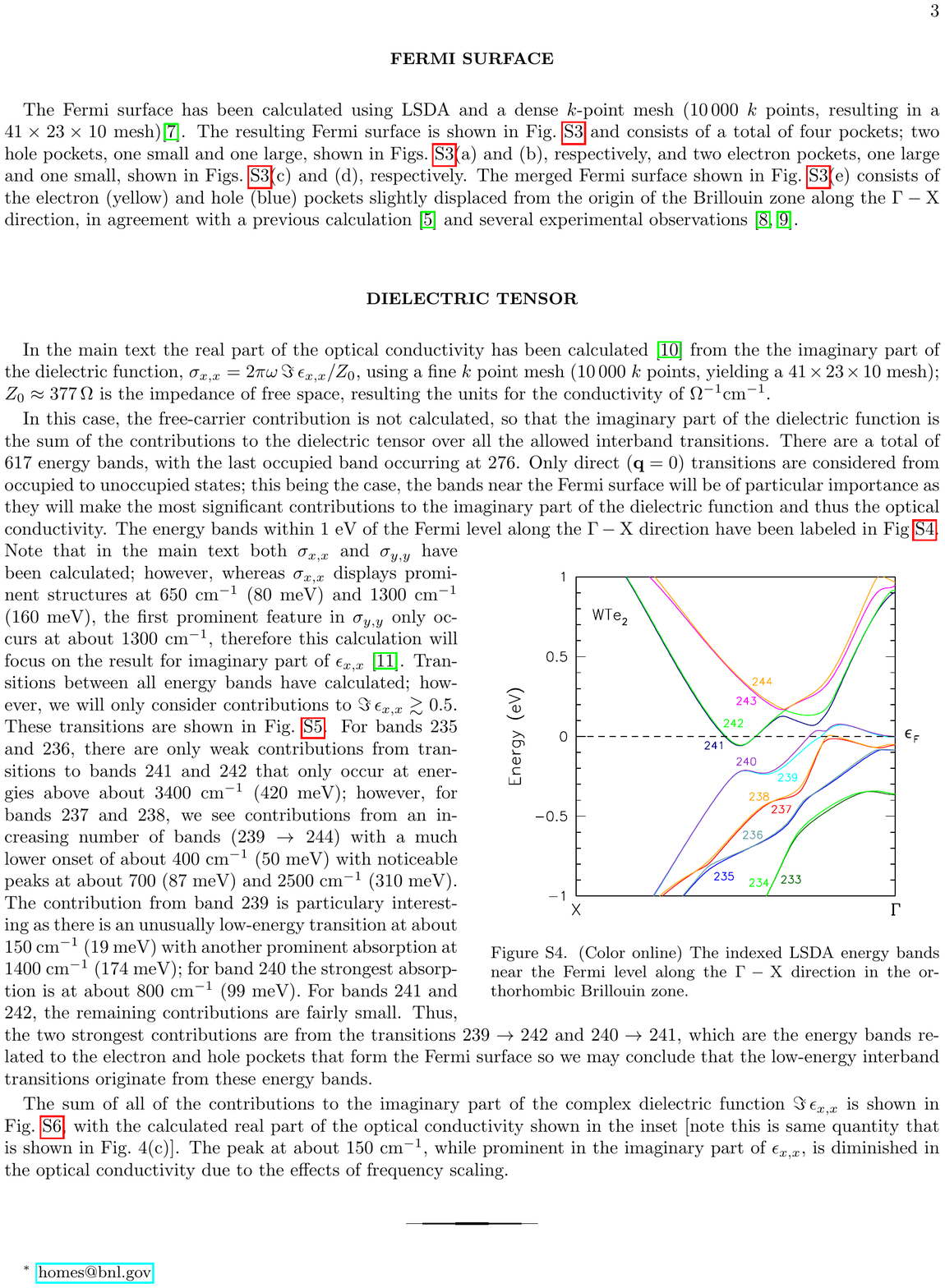} \ \\ % flush the page

\newpage
\includepdf{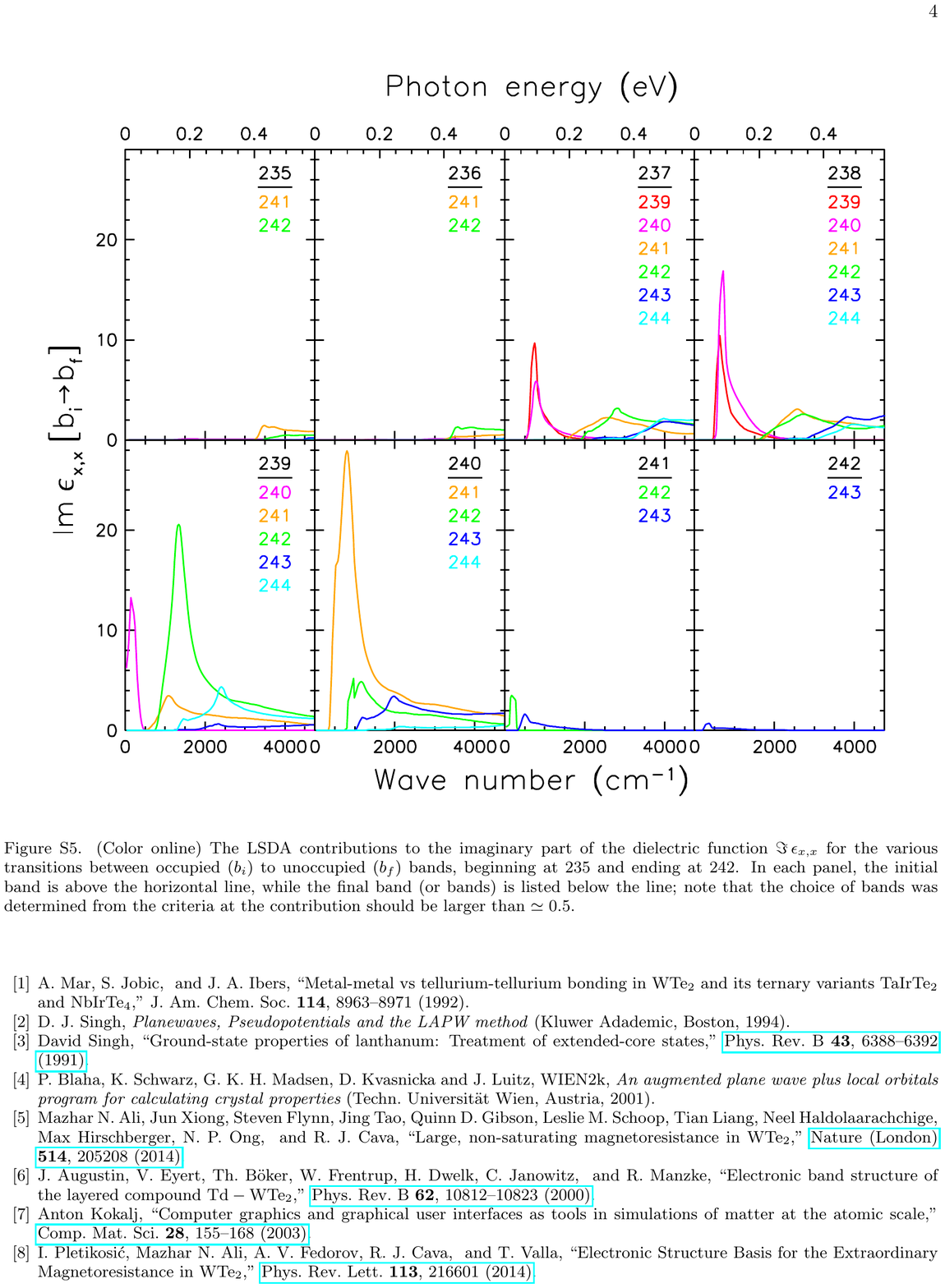} \ \\ % flush the page

\newpage
\includepdf{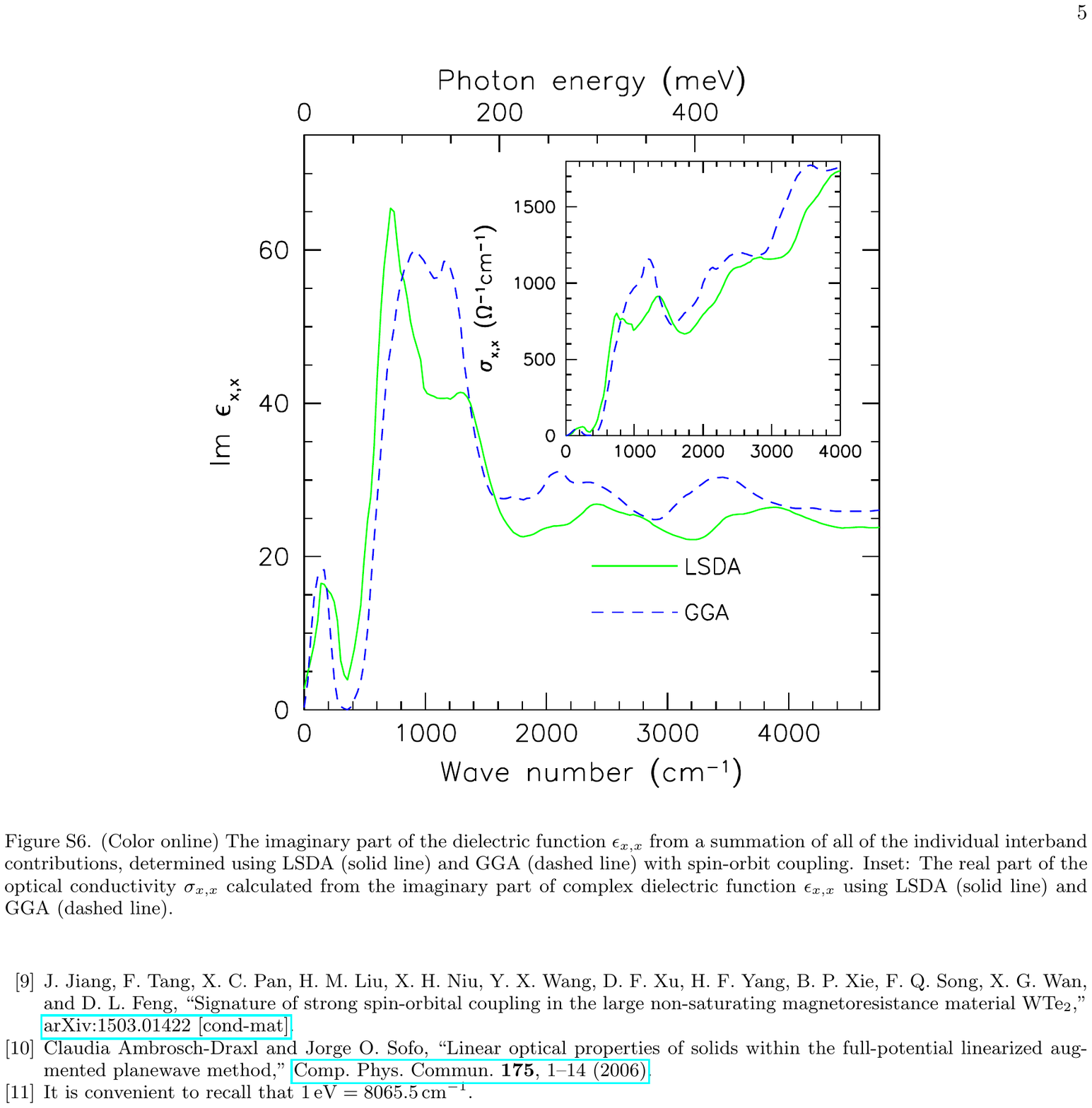} \ \\ % flush the page

\end{document}